\documentclass[aps,prl,twocolumn,floatfix,superscriptaddress,nofootinbib]{revtex4-2}
\usepackage[dvipdfmx]{graphicx}
\usepackage{bm,amsmath,amssymb,cases,braket}
\usepackage{times}
\usepackage{appendix}
\usepackage{lineno}

\usepackage[normalem]{ulem}
\usepackage{color}

\begin{document}

\title{Classical simulation and theory of quantum annealing in a thermal environment}
\author{Hiroki Oshiyama}
\email{hiroki.oshiyama.e6@tohoku.ac.jp}
\affiliation{Department of Physics, Tohoku University, Sendai 980-8578, Japan}
\altaffiliation[Present address: ]{Graduate School of Information Sciences, Tohoku University, Sendai 980-8578, Japan}
\author{Sei Suzuki}
\email{sei01@saitama-med.ac.jp}
\affiliation{Department of Liberal Arts, Saitama Medical University, Moroyama, Saitama 350-0495, Japan}
\author{Naokazu Shibata}
\email{shibata@cmpt.phys.tohoku.ac.jp}
\affiliation{Department of Physics, Tohoku University, Sendai 980-8578, Japan}
\begin{abstract}
We study quantum annealing in the quantum Ising model coupled to a thermal environment. When the speed of quantum annealing is sufficiently slow, the system evolves following the instantaneous thermal equilibrium. This quasistatic and isothermal evolution, however, fails near the end of annealing because the relaxation time grows infinitely, therefore yielding excess energy from the thermal equilibrium. We develop a phenomenological theory based on this picture and derive a scaling relation of the excess energy after annealing.
The theoretical results are numerically confirmed using a novel non-Markovian method that we recently proposed based on a path-integral representation of the reduced density matrix and the infinite time evolving block decimation. In addition, we discuss crossovers {from weak to strong coupling as well as from the adiabatic to quasistatic regime}, and propose experiments on the D-Wave quantum annealer.
\end{abstract}
\maketitle
\emph{Introduction.}---
The quantum annealing (QA) device manufactured by D-Wave
Systems has made an immense impact not only in the physics community
but also in the industrial community with a hope of developing quantum computers and
simulators \cite{Johnson2011, boixo2014, denchev2016, albash2018, King2018, Harris2018, king2019scaling, ushijima-mwesigwa2017, teplukhin2020}. It is known that this device carries out QA
imperfectly in the sense that the system embedded in this
device is affected by its environment \cite{Johnson2011, boixo2013,  Marshall2017}. This fact raises issues regarding
QA dynamics in a thermal environment \cite{sarandy2005, Amin2008, dickson2013, boixo2014, Amin2015, arceci2017, Smelyanskiy2017, smirnov2018}.

QA was proposed as a quantum mechanical algorithm to solve combinatorial optimization problems \cite{Kadowaki1998, Finnila1994, das2005, das2008, albash_RMP2018, hauke2020}. 
The problem to be solved is encoded in an Ising Hamiltonian such that the solution is given by its ground state. The original algorithm is based on the quantum adiabatic time evolution from a known trivial ground state of an initial Hamiltonian to the unknown ground state of the Ising Hamiltonian \cite{farhi2001}. 
However, a system in a quantum device cannot be free from environmental effects. Studying QA in a thermal environment is beneficial not only for QA devices but also to understand nonequilibrium statistical mechanics \cite{patane2008,patane2009,yin2014,hwang2015,nalbach2015,dutta2016,keck2017,arceci2017,weinberg2020,puebla2020,rossini2020,Bando2020,bandyopadhyay2020}.

A plausible picture of QA in the presence of a thermal environment is quasistatic and isothermal evolution, in which a system evolves maintaining a thermal equilibrium state at the temperature of its environment. This picture should be valid when the QA duration is much longer than the relaxation time of the system. Previous studies, based {on} a system-bath coupling realistic in the D-Wave quantum annealer \cite{Albash2012, Amin2015, Marshall2017, Marshall2019}, have suggested that the relaxation time increases dramatically as the transverse field is reduced. Because of this increase, the quasistatic and isothermal evolution should fail near the end of annealing, and result in a final state with an effective temperature higher than that of the environment. Even though this picture has only been studied in small-sized systems, the scalings of the physical quantities expected in the thermodynamic limit have not yet been studied.
In this letter, we develop a phenomenological theory 
and derive a novel scaling relation of the energy after QA.

To study the QA of a system with an experimentally 
realistic system-bath coupling, we employ a
novel numerical non-Markovian method proposed by the present authors in Ref. \cite{Oshiyama2020}. This method makes use of the discrete-time path integral for a dissipative system \cite{makri1992}
and the infinite time-evolving block decimation (iTEBD) algorithm \cite{Orus2008}, which
enable the computation of the reduced density matrix (RDM) in and out of equilibrium of the translationally invariant quantum Ising chain in the thermodynamic limit. We verify the theoretical consequences on QA in a thermal environment using this method.

\emph{Model.}---
We consider the dissipative quantum Ising chain (DQIC) described by the 
Hamiltonian $H(s)=H_{\mathrm{S}}(s)+H_{\mathrm{B}}+H_{\mathrm{SB}}$, 
where $H_{\mathrm{S}}(s)$ represents the system Hamiltonian given by the quantum Ising chain,
\begin{equation}
H_{\mathrm{S}}(s) = A(s) H_{\rm TF}
 + B(s) H_{\rm I} ,
\end{equation}
with $H_{\rm TF} = -\sum_{j=1}^N\hat\sigma_j^x$ and
$H_{\rm I} = - \sum_{j=1}^{N-1}\hat{\sigma}_j^z\hat{\sigma}_{j+1}^z$.
Here $\hat\sigma^x_{j}$ and $\hat\sigma^z_{j}$ denote the Pauli matrices for the site $j$,
$N$ is the number of sites, and $s$ is a parameter ranging from 0 to 1. 
The schedule functions, $A(s)$ and $B(s)$, are assumed to be
\begin{equation}
A(s)=(1-s)^\alpha,~~B(s)=s, \label{schedule}
\end{equation}
where an exponent $\alpha>0$ in $A(s)$ represents how the transverse field goes to zero at the end of annealing. The bath Hamiltonian is represented by the collection of harmonic oscillators,
$H_{\mathrm{B}}=\sum_{j=1}^{N}\sum_{k}\omega_{k}\hat b^{\dagger}_{j,k}\hat b_{j,k}$, 
where $\hat b_{j,k}$ and $\hat b_{j,k}^{\dagger}$ are the annihilation and creation operators, respectively, of the boson for the site $j$ and mode $k$ with the frequency $\omega_k$. We use the unit $\hbar = 1$
throughout this letter.
As for the interaction between the system and the bath, we assume the Caldeira--Leggett model \cite{Caldeira1983,Leggett1987} for dissipative superconductor flux qubits given by
\begin{equation}
H_{\mathrm{SB}}=\sum_{j=1}^{N}\hat\sigma^z_{j}\sum_{k}
\lambda_{k}\left(\hat b^{\dagger}_{j,k}+\hat b_{j,k}\right),
\label{HSB}
\end{equation}
where $\lambda_{k}$ is a coupling constant for the mode $k$. 
We assume the Ohmic spectral density for the bath modes, 
\begin{equation}
J(\omega)=\sum_{k}\lambda_{k}^2\delta(\omega-\omega_{k})=
\frac{\eta}{2}\omega e^{-\omega/\omega_c},
\end{equation}
where $\eta$ is the dimensionless 
coupling constant and $\omega_c$ is the cut-off frequency of {the bath spectrum}, which is chosen to be larger than the bath temperature.
We leave quantitative study for {other system-bath couplings and} a non-Ohmic bath to future work.
When we mention QA, we consider the time evolution with time $t$ from
$t = 0$ to $t_{\rm a}$ by the Hamiltonian $H(t/t_{\rm a})$. 

The dynamics of the spin system is specified by the RDM defined by tracing out the bosonic degrees of freedom from the density matrix $\rho(t)$ of the full system, 
\begin{equation}
\rho_{\mathrm{S}}(t)\equiv\mathrm{Tr}_{\mathrm{B}}\rho(t)
=
\mathrm{Tr}_{\mathrm{B}}\left[
\mathcal{U}(t)\rho(0)
\mathcal{U}^{\dagger}(t)
\right] ,
\label{rdm}
\end{equation}
where $\mathrm{Tr}_{\rm B (S)}$ stands for the trace with respect to the boson (spin) degrees of freedom, $\mathcal{U}(t)$ is the time evolution operator of the full system and $\rho(0)$ is an initial density matrix. We assume that $\rho(0)$ is the direct product of the ground state of $H_{\mathrm{S}}(0)$ denoted by $\ket{\psi_0(0)}$ and the thermal equilibrium state of $H_{\mathrm{B}}$ at the temperature $T_{\rm B}$:
$\rho(0)=\ket{\psi_0(0)}\bra{\psi_0(0)}\otimes 
{e^{-H_{\mathrm{B}}/T_{\rm B}}}/{Z_{\mathrm{B}}}$,
where $Z_{\mathrm{B}}$ is the partition function of $H_{\mathrm{B}}$. We refer to $T_{\rm B}$ as the bath temperature. We {choose} the Boltzmann constant $k_B$ to be the temperature unit throughout this letter. 

The spin state in the instantaneous thermal equilibrium at $s$ and temperature $T$ is given by
\begin{align}
\rho_{\mathrm{S}}^{\mathrm{eq}}(s, T)\equiv{\mathrm{Tr}_{\mathrm{B}}[e^{-H(s)/T}}]/{Z(s, T)},
\label{eqrdm}
\end{align}
where $Z(s, T)$ is the partition function of the full system.
We define the Gibbs state of $H_{\rm I}$ as
\begin{equation}
 \rho_{\rm I}^{\mathrm{eq}}(T) \equiv e^{-H_{\rm I}/T}/{\rm Tr}_{\rm S}[ e^{-H_{\rm I}/T}] .
\end{equation}
Note that $\rho^{\rm eq}_{\rm S}(s, T)$ at $s = 1$ reduces to
$\rho_{\rm I}^{\rm eq}(T)$ 
because $H_{\rm S}(1)$ commutes with $H_{\rm SB}$ and 
${\rm Tr}_{\rm B}e^{- (H_{\rm S} + H_{\rm SB})/T}$ is independent of
$\sigma_j^z$, which is shown by introducing new boson operators 
$\tilde{b}_{j,k}\equiv b_{j,k} + \lambda_k\sigma_j^z/\omega_k$ for
all $j$ and $k$.

\emph{Non-Markovian iTEBD.}--- We focus on a time-dependent state
and outline the numerical method \cite{Oshiyama2020} used to compute Eq.~(\ref{rdm}). 
The application to the equilibrium RDM in Eq.~(\ref{eqrdm}) is straightforward.

Let us apply the Trotter decomposition \cite{Trotter1959, Suzuki1976} with a step size $\Delta t = t/M$ and the Trotter number $M$ to $\mathcal{U}(t)$ in Eq.~(\ref{rdm}), and perform the Gaussian integral with respect to the bosonic degrees of freedom. The resulting discrete-time path integral formula of the RDM is given by
\begin{align}
\bra{\boldsymbol{\sigma}^{(M)}}\rho_{\mathrm{S}}(t)\ket{\boldsymbol{\sigma}^{(M+1)}}
 =\sum_{\{\sigma_j^{(l)}=\pm 1\}_{l\neq M,M+1}} e^{i\mathcal{S}_{0}+\mathcal{S}_{\mathrm{infl}}} ,
\label{QUAPI}
\end{align}
where $\sigma_{j}^{(l)}$ is the Ising-spin variable at the site $j$ and the time $t_l$ is defined as
\begin{align}
t_l=\left\{
\begin{array}{cl}
l\Delta t & (0\leq l \leq M)\\
(2M+1-l)\Delta t & (M+1 \leq l \leq 2M+1), 
\end{array}
\right.
\end{align}
and $\ket{\boldsymbol{\sigma}^{(l)}}$ denotes the eigenstate of $\sigma_{j}^z$
with the eigenvalue $\sigma_j^{(l)}$ \cite{Caldeira1983, Makarov1994}. 
$\mathcal{S}_0$ denotes the action of the isolated spin system.

The influence action $\mathcal{S}_{\rm infl}$ induced by coupling to the bath is given by
\begin{align}
\mathcal{S}_{\mathrm{infl}}=\sum_{j=1}^{N}\sum_{l>m}^{|t_l-t_m|<\tau_c}\kappa_{l,m}\sigma_{j}^{(l)}\sigma_{j}^{(m)} ,
\label{infl}
\end{align}
where
\begin{equation}
\kappa_{l,m} = \Delta t^2 
\int_0^{\infty}d\omega J(\omega)\frac{\cosh[\omega/(2T_{\rm B})-i\omega(t_l - t_m)]}{\sinh[\omega/(2T_{\rm B})]}.
\label{kernel} 
\end{equation}
Note that $\tau_c$ in Eq. (\ref{infl}) is the memory time cut-off introduced to reduce the computational cost. 

\begin{figure*}[t]
\begin{center}
\includegraphics[width=2\columnwidth, height=.45\columnwidth]{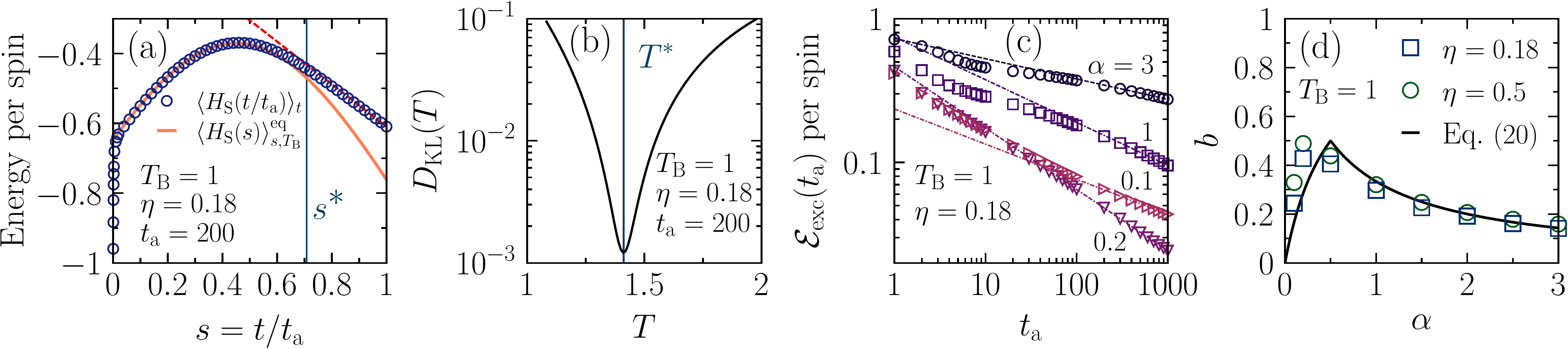}
\end{center}
\caption{(a) Energy expectation values, $\langle H_{\rm S}(t/{t_{\rm a}})\rangle_t$ 
and $\langle H_{\rm S}(s)\rangle_{s,T_{\rm B}}^{\rm eq}$, per spin of the time-dependent state $\rho_{\rm S}(t)$ and the instantaneous thermal equilibrium state $\rho_{\rm S}^{\rm eq}(s, T_{\rm B})$, respectively, as functions of the rescaled time $s$ for $t_{\mathrm{a}}=200$ and $\eta=0.18$ at $T_{\rm B}=1$.  We fixed $\alpha=1$. 
The dashed line and the solid vertical line indicate $\frac{1}{N}\langle H_{\rm I}\rangle_{t_a}t/t_a$ and $s^* \equiv T_{\rm B}/T^{\ast}$, respectively, where $T^{\ast}$ is determined by the minimization of the Kullback--Leibler (KL) divergence.
(b) KL divergence $D_{\rm KL}(T)$ between the final state after QA and the Gibbs state of $H_{\rm I}$ with temperature $T$. See the main text for a detailed definition. 
(c) Excess energy $\mathcal{E}_{\mathrm{exc}}$ per spin from the thermal expectation value after QA as a function of $t_{\mathrm{a}}$ for various $\alpha$. Lines indicate the best power-law fits $\mathcal{E}_{\rm exc}=at_a^{-b}$ to the data for $t_{\mathrm{a}}>100$ with the fitting parameters $a$ and $b$. 
(d) Exponent $b$ as a function of $\alpha$. The numerical results (symbols) are compared to the theoretical prediction shown by the solid line. The parameters used in the numerical simulations are $\omega_c=5$, $\tau_c=10$, $\Delta t=0.05$ and $N\to\infty$. The bond dimensions are up to 128.} 
\label{fig:fT}
\end{figure*}

The key idea of our method is to represent the part of $\exp \mathcal{S}_{\mathrm{infl}}$ associated with a site $j$ in terms of a matrix product state (MPS) as follows: 
\begin{align}
&\exp\Biggl(\sum_{l>m}^{|t_l-t_m|<\tau_c}\kappa_{l,m}\sigma_{j}^{(l)}\sigma_{j}^{(m)}\Biggr)
\nonumber \\
&
\approx \sum_{\{\mu_{j,l}\}}^{\chi_t}
\phi^{(j,0)S^{(0)}_{j}}_{\mu_{j,0}}
\phi^{(j,1)S^{(1)}_{j}}_{\mu_{j,0},\mu_{j,1}}
\phi^{(j,2)S^{(2)}_{j}}_{\mu_{j,1},\mu_{j,2}}\cdots\phi^{(j,M)S^{(M)}_{j}}_{\mu_{j,M-1}},
\label{infl_mps}
\end{align}
where $S_j^{(l)}\equiv (\sigma_{j}^{(l)},~\sigma_{j}^{(2M+1-l)})$ denotes the composite variable and $\chi_t$ is the bond dimension which controls the precision of the approximation in this MPS representation. {The tensors $\phi^{(j,l)}$} {are given by recursive application of the singular value decomposition \cite{Suzuki2019,Oshiyama2020}}. Using Eq. (\ref{infl_mps}) in Eq. (\ref{QUAPI}), we obtain a tensor network representation for the RDM:
\begin{align}
&\bra{\boldsymbol{\sigma}^{(M)}}\rho_{\mathrm{S}}(t)\ket{\boldsymbol{\sigma}^{(M+1)}} \\
&\approx\sum_{\{S_{j}^{(l)},\mu_{j,l}\}_{l\neq M}}
e^{i\mathcal{S}_{\mathrm{0}}}
\prod_{j=1}^{N}\phi^{(j,0)S^{(0)}}_{\mu_{j,0}}
 \Bigl[\prod_{l=1}^{M-1}\phi_{\mu_{j,l-1},\mu_{j,l}}^{(j,l)S_{j}^{(l)}}\Bigr]
\phi^{(j,M)S^{(M)}_{j}}_{\mu_{j,M-1}} .\nonumber 
\end{align}
Having obtained this tensor network representation, the iTEBD algorithm can be applied to implement the sum
with respect to $\{S_{j}^{(l)},\mu_{j,l}\}_{l\neq M}$ and compute local quantities, taking $N\to\infty$ and using the translational invariance in space \cite{Orus2008}.

\emph{Phenomenological theory.}--- Let us assume a
finite bath temperature $T_{\rm B} > 0$. 
In the limiting case of $t_{\rm a}\to\infty$,
QA in this thermal environment leads to the quasistatic and isothermal
process. Accordingly, the final state of the spin system is described by 
$\rho^{\rm eq}_{\rm S}(1, T_{\rm B}) = \rho_{\rm I}^{\rm eq}(T_{\rm B})$. 
When $t_{\rm a}$ is finite, the spin system approximately maintains thermal equilibrium as long as
the relaxation time of the spin system is shorter than the annealing time scale.
However, {in the case of $[H_{\rm S}(1),H_{\rm SB}]= 0$, it is known that the relaxation time grows infinitely with $s\to 1$. Therefore the quasistatic and isothermal evolution must fail before
QA ends, and the spin state is expected to be frozen at a time $t^{\ast} = s^{\ast}t_{\rm a}$ \cite{Albash2012,Amin2015}}. 
We refer to $t^{\ast}$ or $s^{\ast}$ as the freezing time. To develop a scaling theory for the freezing time, we employ the quasistatic-freezing
approximation as follows.
The quasistatic-freezing approximation assumes that the spin state is frozen when the changing rate with $t$ of the instantaneous relaxation time of the spin system exceeds unity.
Writing the instantaneous relaxation time at $s = t/t_{\rm a}$ as $\tau_{\mathrm{rel}}(s)$, 
the freezing time is then determined by
\begin{align}
\dot{\tau}_{\mathrm{rel}}(s^*)=1 ,
\label{s_eq}
\end{align}
where the dot denotes differentiation by $t$. $\tau_{\mathrm{rel}}(s)$ is now estimated from the transition rate $\gamma(s)$. Using 
Fermi's golden rule, the latter is given, up to an $s$-independent factor, as
\begin{align}
\gamma_{l,m}(s)\propto \eta|\bra{\psi_l(s)}\sum_{i}\hat\sigma^z_{i}
\ket{\psi_m(s)}|^2
\label{rel}
\end{align}
where $\ket{\psi_l(s)}$ and $\ket{\psi_m(s)}$ are the $l$-th and $m$-th eigenstates of 
$H_{\mathrm{S}}(s)$, respectively. 
{When $s$ is close to 1, 
$|\psi_l(s)\rangle$ is written within the first order of $A(s)$ as $|\psi_l(s)\rangle\approx|\psi_l(1)\rangle + A(s)\sum_{m\neq l}[\langle\psi_l(1)|H_{\rm TF}|\psi_m(1)\rangle/(E_l-E_m)]|\psi_m(1)\rangle$, where $E_m$ denotes an eigenenergy of $B(s)H_{\rm I}$. Using this and noting that $\sum_i\hat\sigma^z_{i}$ is diagonal with the basis $\{|\psi_m(1)\rangle\}$, one finds that Eq.~(\ref{rel}) is proportinal to $\eta A(s)^2$. Therefore, the scaling of relaxation time is obtained as}
\begin{align}
\tau_{\mathrm{rel}}(s)\approx\gamma_{l,m}(s)^{-1}\sim \eta^{-1}A(s)^{-2} \sim \eta^{-1}(1-s)^{-2\alpha} .
\label{rel_res}
\end{align}
Using this in Eq. (\ref{s_eq}), the scaling relation of $s^{\ast}$ is obtained as follows:
\begin{align}
(1-s^{\ast})\sim (\eta t_{\mathrm{a}})^{-1/(2\alpha+1)}. 
\label{t_frz}
\end{align}
Now, the quasistatic-freezing approximation implies that the RDM
after the freezing time is approximately replaced by that of the instantaneous thermal equilibrium at $s = s^{\ast}$,
namely, $\rho_{\rm S}(t)\approx \rho_{\rm S}^{\rm eq}(s^{\ast}, T_{\rm B})$ for $t > s^{\ast}t_{\rm a}$. 
Moreover, $\rho_{\rm S}^{\rm eq}(s^{\ast}, T_{\rm B})$ {can be approximated by the Gibbs state $\rho_{\rm S}^{\rm Gibbs}\equiv e^{-H_{\rm S}(s^{\ast})/T_{\rm B}}/{\rm Tr}[e^{-H_{\rm S}(s^{\ast})/T_{\rm B}}]$ for sufficiently weak $\eta$, and the latter is approximated as 
$\rho_{\rm S}^{\rm Gibbs}\approx \rho_{\rm I}^{\rm eq}(T_{\rm B}/{B(s^{\ast})}) + 
O(A(s^{\ast}))$ near $s^{\ast}=1$.} 
Therefore, neglecting the {$O(A(s^{\ast})^2)$ and $O(A(s)A(s^{\ast}))$} terms 
for $\alpha > \frac{1}{2}$, the energy of the spin system for $t > s^{\ast}t_{\rm a}$ is estimated {as \cite{Note_eq18}}
\begin{equation}
 \langle H_{\rm S}(t/t_{\rm a}) \rangle_t \approx 
 B(t/t_{\rm a})\langle H_{\rm I}\rangle^{\rm eq}_{{\rm I}, T_{\rm B}/{B(s^{\ast})}} ,
\label{sst}
\end{equation}
where $\langle \cdot\rangle_t$ and
$\langle \cdot\rangle^{\rm eq}_{{\rm I},T}$ represent the expectation values
with respect to $\rho_{\rm S}(t)$ and $\rho_{\rm I}^{\rm eq}(T)$, respectively.
Therefore, the energy of the spin system
approaches the thermal expectation value
of $H_{\rm I}$ at the temperature $T_{\rm B}/{B(s^{\ast})}$ 
as $s\to 1$.
For general $\alpha > 0$,
{expanding $\rho_{\rm S}^{\rm Gibbs}$ in series of $(1-s^{\ast})$ and $(1-s^{\ast})^{\alpha}$ perturbatively, one obtains $\langle H_{\rm S}(1)\rangle_{t_{\rm a}}
\approx \langle H_{\rm S}(1)\rangle^{\rm eq}_{{\rm I}, T_{\rm B}} + 
c_1 (1 - s^{\ast})+c_2 (1 - s^{\ast})^{2\alpha}$ \cite{Note_eq18}, where $c_1$ and $c_2$ are coefficients independent of $s^{\ast}$. 
}
{Keeping the leading term and} applying Eq. (\ref{t_frz}), the excess energy of the final state is obtained as
\begin{align}
\mathcal{E}_{\rm exc} \equiv \langle H_{\rm S}(1)\rangle_{t_{\rm a}}
 - \langle H_{\rm S}(1)\rangle^{\rm eq}_{{\rm I}, T_{\rm B}}
\sim (\eta~ t_{\mathrm{a}})^{-b} ,
\label{Eexc_pow}
\end{align}
with
\begin{align}
 b = \min\{1,2\alpha\}/({2}\alpha + 1) .
\label{Eexc_pow_b}
\end{align}
Note that the excess energy decays the fastest when $\alpha = \frac{1}{2}$.
Equations (\ref{Eexc_pow}) and (\ref{Eexc_pow_b}) are valid for DQIC in any dimension, any lattice and non-Ohmic spectral densities as well.

\emph{Numerical results.}--- 
Figure \ref{fig:fT}(a) shows the energy expectation value
per site of the time-dependent state during QA and that of the instantaneous thermal equilibrium
as functions of the rescaled time $s$. After the initial relaxation, 
the system maintains thermal equilibrium until a certain time $s^{\ast}$, when the quasistatic and isotheral evolution fails and the energy deviates upwards from that of the instantaneous equilibrium state. This behavior 
is perfectly consistent with the quasistatic-freezing picture
mentioned above. To evaluate the freezing time $s^{\ast}$ and identify the final energy
$\langle H_{\rm I}\rangle_{{t_{\rm a}}}$, 
we focus on the Kullback-Leibler (KL) divergence $D_{\rm KL}$ of the final
state and the Boltzmann distribution of $H_{\rm I}$ as
a measure of the distance between the two.
Because this quantity is not accessible for the RDMs of the entire spin system when using our method,
we instead consider the RDMs of eight spins given by
$\rho_{8} \equiv {\rm Tr}_{\bar{8}}\rho_{\rm S}({t_{\rm a}})$ and
$\rho_{8}^{\rm eq}(T)\equiv {\rm Tr}_{\bar{8}}\rho_{\rm I}^{\rm eq}(T)$
to define $D_{\rm KL}(T) \equiv {\rm Tr}_{8}
[\rho_8(\log\rho_8 - \log\rho_8^{\rm eq}(T))]$, where
${\rm Tr}_{8}$ and ${\rm Tr}_{\bar{8}}$ denote the trace with respect to 
the eight adjacent spins and the other spins, respectively.
We show $D_{\rm KL}(T)$ in Fig.~\ref{fig:fT}(b). $D_{\rm KL}(T)$
has a sharp minimum at a certain $T$ labeled $T^{\ast}$. This implies that
the RDM after QA is approximated by the Gibbs state of $H_{\rm I}$ with the temperature $T^{\ast}$. In addition, as shown in Fig.~\ref{fig:fT}(a),
the curve of $\langle H_{\rm S}(t/{t_{\rm a}})\rangle_t$ is indistinguishable from the line of 
$
s\langle H_{\rm I}\rangle_{{\rm I}, T^{\ast}}$ near $s = 1$. Assuming $T^{\ast} = T/
s^{\ast}$,
this result implies Eq.~(\ref{sst}) and that
$s^{\ast}$ determined by $T/T^{\ast}$ is consistent with the freezing time when the quasistatic evolution fails (see the vertical line in Fig.~\ref{fig:fT}(a)).
Figure \ref{fig:fT}(c) shows the excess energy as a function of the annealing time $t_{\rm a}$ for $\eta = 0.18$ and $T_{\rm B} = 1$. It can be seen that the excess energy decays as a power law for large $t_{\rm a}$ with an exponent denoted by $b$ that depends on $\alpha$. Figure \ref{fig:fT}(d) shows the $\alpha$ dependence of the exponent $b$. There is excellent
agreement between the numerical results and the theoretical prediction.

\begin{figure}[t]
\begin{center}
 \includegraphics[width=0.8\columnwidth, height=0.6\columnwidth]{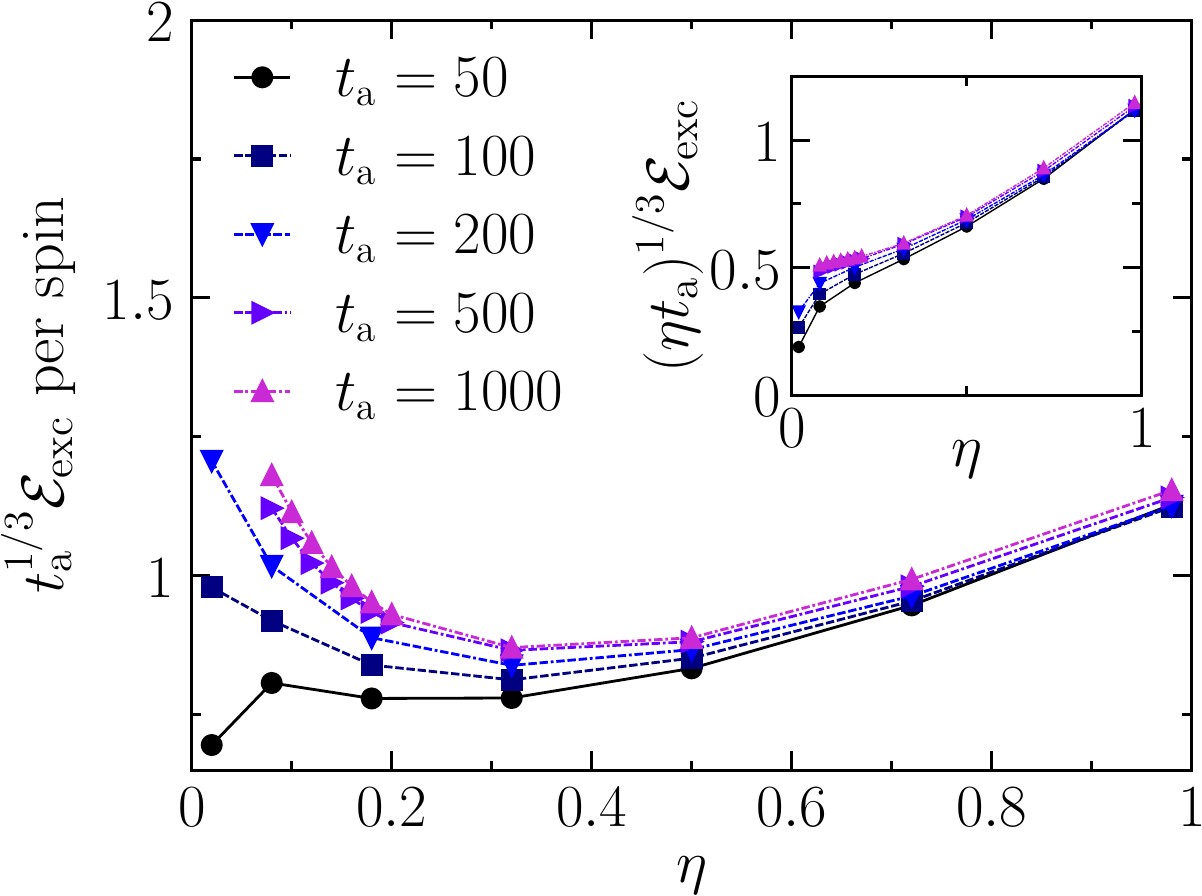}
\end{center}
\caption{Excess energy per spin after QA scaled by $t_\mathrm{a}^{-1/3}$ as a function of $\eta$ for $T_{\rm B} = 1$, $\alpha = 1$, and
various ${t_{\rm a}}$ ranging from 50 to 1000. With increasing $t_{\rm a}$, the data collapse into a single nonmonotonic curve, which implies $\mathcal{E}_{\rm exc}\sim t_{\rm a}^{-1/3}$ for large $t_{\rm a}$.
The inset shows the data rescaled by $(\eta t_\mathrm{a})^{-1/3}$.
The constancy of the data with large $t_{\rm a}$ near $\eta = 0$ corresponds to Eq.~(\ref{Eexc_pow}).
The parameters in the numerical simulations are the same as those in Fig.~\ref{fig:fT}.}
\label{fig:E-eta}
\end{figure}

Figure \ref{fig:E-eta} shows the $\eta$-dependence of $\mathcal{E}_{\rm exc}$ scaled by $t_\mathrm{a}^{1/3}$ for 
$T_{\rm B} = 1$, $\alpha = 1$, and various ${t_{\rm a}}$. 
It can be seen that $\mathcal{E}_{\rm exc}$ is nonmonotonic with respect to $\eta$.
The decreasing behavior of $\mathcal{E}_{\rm exc}$ with increasing 
$\eta$ in the weak coupling regime is consistent with Eq.~(\ref{Eexc_pow}), 
while its increasing
behavior in the strong-coupling regime ($\eta \gtrsim 0.4$) is not described by the phenomenological theory mentioned above. This failure of the theory arises from the perturbative argument used for the relaxation time in Eq.~(\ref{rel}).
The existence of the optimal strength in the system-bath coupling to reduce $\mathcal{E_{\rm exc}}$ is
first revealed by our numerical method based on a non-perturbative formulation. Note that the scaling of $\mathcal{E}_{\rm exc}$ by $t_{\rm a}$ is valid even in the case of strong-coupling.

\begin{figure}[t]
\begin{center}
\includegraphics[width=0.8\columnwidth, height=0.6\columnwidth]{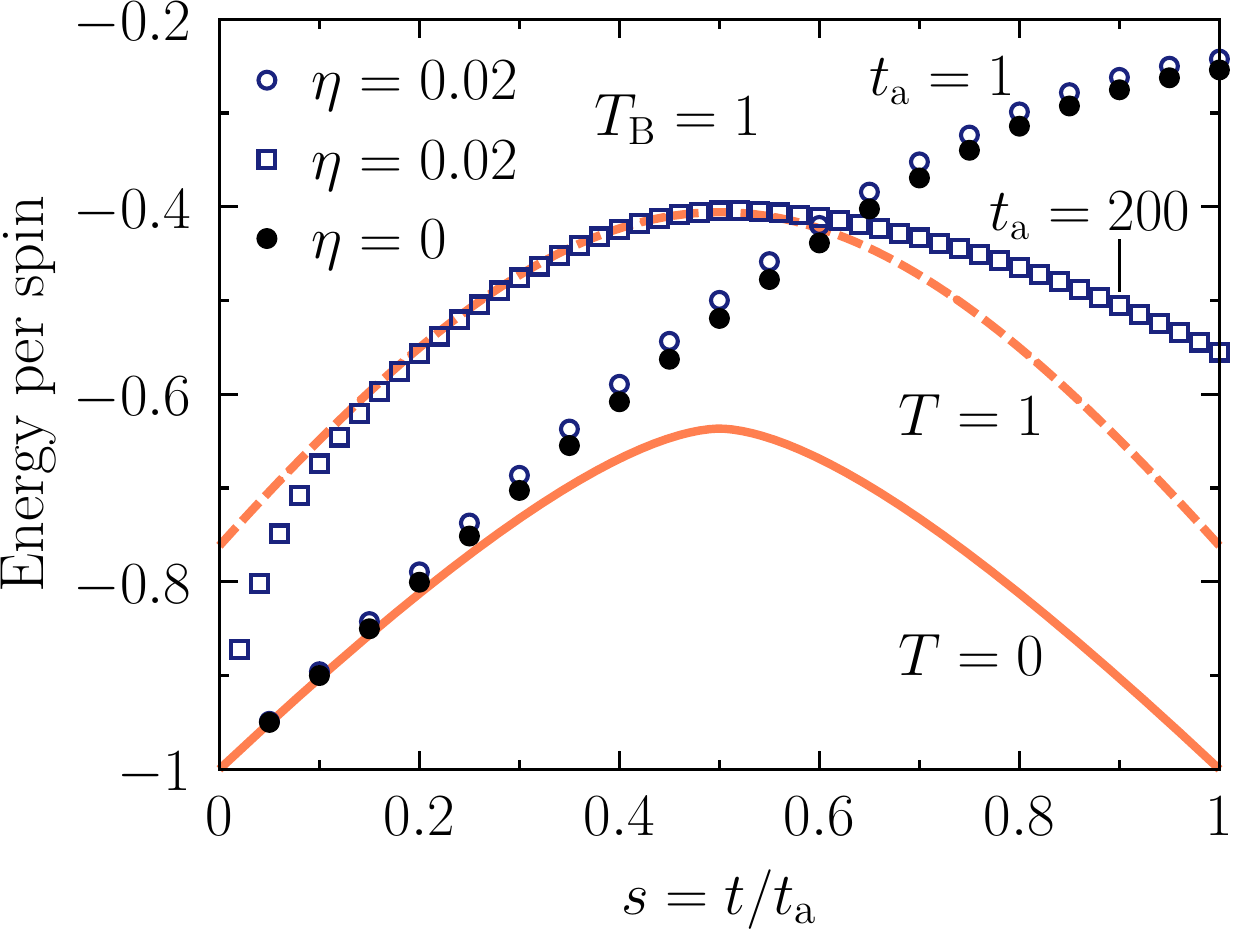}
\end{center}
\caption{
Energy expectation values per spin as functions of the rescaled time $s$.
{The solid and dashed lines show the energies per spin of the instantaneous equilibrium states of the closed system at $T = 0$ and $T = 1$ }{, respectively}. 
The filled symbols show the energy of the time-dependent state for $t_{\rm a} = 1$
of the closed system ($\eta=0$).
The empty symbols denote energies of the time-dependent states
of the dissipative system 
with $\eta = 0.02$ and $T_{\rm B} = 1$
for $t_{\rm a} = 1$ and $t_{\rm a} = 200$.
We fixed $\alpha = 1$. 
The parameters in the numerical simulations are the same as those in Fig.~\ref{fig:fT}.}
\label{fig:small_eta}
\end{figure}

So far, we have focused on slow QA in a thermal environment with a finite temperature and have discussed consequences of freezing near the end of annealing.
Here, we comment on two situations where the dynamics is governed by a quantum phase transition {(QPT)} at zero temperature, assuming the absence of a thermal phase transition at finite temperature. 
The first is the case of weak-coupling and short-annealing-time. When the system-bath coupling is sufficiently weak, i.e. $\eta\ll 1$, QA drives the spin system in the same way as a closed system as long as ${t_{\rm a}}$ is not large, as demonstrated by the {proximity of} {filled and empty} circles in Fig.~\ref{fig:small_eta}. 
{In this case, the QPT governs the dynamics, and} gives rise to the Kibble--Zurek scaling (KZS) \cite{Kibble1976, Zurek1985, Dziarmaga2005} of the residual energy to the ground state after QA.
For larger ${t_{\rm a}}$, the system is thermalized 
{and the QPT no longer affect the dynamics} 
as shown by the {overlap of} squares {with the dashed line} in Fig.~\ref{fig:small_eta}. {The crossover from the KZS regime to the large $t_{\rm a}$ regime} is accompanied by a non-monotonic change in the residual energy {when $T_{\rm B}$ is sufficiently high} \cite{SM}. The second is the case of medium-coupling and low-temperature, where the dynamics is governed by a {QPT} of the dissipative system at {$T_{\rm B}=0$}. In this case, KZS with a modified exponent \cite{Oshiyama2020} is observed. When the temperature is not low and/or ${t_{\rm a}}$ is much larger, however, the spin system is not influenced by a {QPT} and the quasistatic-freezing picture is valid because the time scale of QA is beyond the characteristic time in the quantum critical region \cite{SM}. 
A recent experimental study suggests that systems realized in the D-Wave device should be in a situation with a medium $\eta$ and a low {$T_{\rm B}$} \cite{Bando2020}.
Therefore, if one performs experiments with still longer ${t_{\rm a}}$ or higher {$T_{\rm B}$}, the scaling of the excess energy given by Eqs.~ (\ref{Eexc_pow}) and (\ref{Eexc_pow_b}) should be observed.

\emph{Summary.}--- {We studied QA in a thermal environment. The simulation using the non-Markovian iTEBD not only confirmed the phenomenological theory for weak system-bath coupling but revealed a nontrivial behavior of the excess energy after QA in the regime beyond weak coupling.} The findings presented here will be beneficial in designing and evaluating QA devices. {Other system-bath couplings, non-Ohmic baths,} and other driven DQICs are open to numerical study with the non-Markovian iTEBD method.

The authors acknowledge H. Nishimori and Y. Susa for valuable discussions, and Y. Bando M. Ohzeki, F. J. G\'omez-Ruiz, A. del Campo, and D. A. Lidar for collaboration on a related experimental project.

\bibliographystyle{apsrev4-1}
\bibliography{DTIM}

\begin{thebibliography}{53}
\bibitem [{Not()}]{Note_eq18}%
  \BibitemOpen
  \href@noop {} {}\bibinfo {note} {{Note here that the terms proportional to
  $A(s)$ vanish in ${\rm Tr}[H_{\rm I}\rho_{\rm S}^{\rm Gibbs}]$, since $H_{\rm
  TF}$ has no diagonal elements in the eigenbasis of $H_{\rm I}$.}}\BibitemShut
  {Stop}%
\end{thebibliography}


\end{document}


\title{Supplementary Meterials: Classical simulation and theory of quantum annealing in a thermal environment}
\author{Hiroki Oshiyama}
\affiliation{Department of Physics, Tohoku University, Sendai 980-8578, Japan}
\author{Sei Suzuki}
\affiliation{Department of Liberal Arts, Saitama Medical University, Moroyama, Saitama 350-0495, Japan}
\author{Naokazu Shibata}
\affiliation{Department of Physics, Tohoku University, Sendai 980-8578, Japan}
\maketitle

In this Supplementary Materials, we supplement numerical results on the residual energy to the ground state after QA, which is defined by $\mathcal{E}_{\rm res}(t_{\rm a})\equiv \langle H_{\rm S}(1)\rangle_{t_{\rm a}} - E_{\rm g}(1)$ where the ground energy $E_{\rm g}(1)$ of $H_{\rm S}(1)$ is given by $E_{\rm g}(1)=-1$.

At first, we briefly review the Kibble-Zurek scaling (KZS) of the residual energy in a closed system. When $\eta = 0$, the spin system encounters a quantum phase transition during QA with the time-dependent Hamiltonian $H_{\rm S}(t)$ changing from $H_{\rm TF}$ to $H_{\rm I}$. The system loses adiabaticity near a quantum critical point and thereby ends up with a residual energy and topological defects. The residual energy is smaller for slower QA, i.e., larger ${t_{\rm a}}$. This mechanism for the residual energy due to time evolution across a continuous phase transition is called as the Kibble-Zurek mechanism (KZM) \cite{Kibble1976, Zurek1985}. The scaling of the residual energy with respect to ${t_{\rm a}}$ is universal and given by $E_{\rm res} \sim {t_{\rm a}}^{-d\nu/({z}\nu + 1)}$ \cite{zurek2005, polkovnikov2005}, where $\nu$ and $z$ are the correlation-length and the dynamical critical exponents of the associated quantum critical point. This scaling of the residual energy is the KZS. In case of the transverse Ising chain, one has $\nu = z = 1$ and hence KZS leads to $E_{\rm res}\sim {t_{\rm a}}^{-1/2}$ \cite{Dziarmaga2005}.

Before proceeding to numerical results, we comment on KZS in the dissipative transverse Ising chain. Consider the case with $\eta > 0$ and $T = 0$. This dissipative spin system also involves a quantum phase transition. The quantum critical point forms a quantum critical line $s_c(\eta)$ in the $s-\eta$ plane. An earlier study using the quantum Monte-Carlo simulation have estimated critical exponents on this critical line as $\nu \approx 0.64$ and $z \approx 2.0$ for $\eta > 0$ \cite{Werner2005}. These exponents suggest a modification of the KZS exponent $b$ of $E_{\rm res}\sim {t_{\rm a}}^{-b}$ as $b_{\rm QMC}\approx 0.28$ from $1/2$ of the closed system. The present authors have studied KZS of the same model by means of iTEBD in the previous work \cite{Oshiyama2020}. The results imply that the exponent of KZS decreases gradually from 0.5 to 0.25 with increasing $\eta$ from 0 to 0.7. The qualitative tendency that the exponent declines in the presence of the environment is shared by the theoretical prediction with quantum Monte-Carlo data and numerical experiments. The quantitative inconsistency is attributed to the smallness of ${t_{\rm a}}$ in numerical experiments. 

\begin{figure}
    \centering
    \includegraphics[width=8cm]{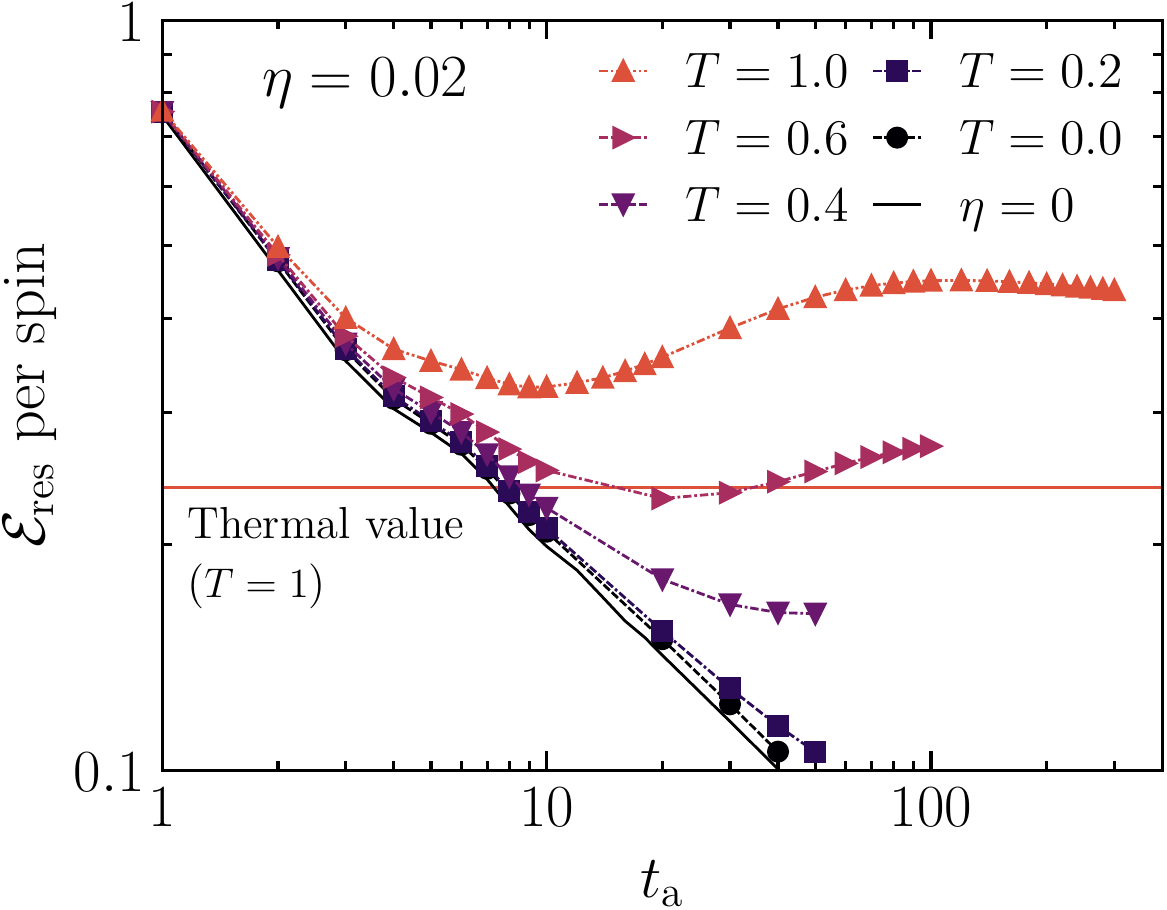}
    \caption{Residual energy after QA as a function of annealing time $t_{\rm a}$ for a weak coupling with $\eta = 0.02$ and various temperatures. The result of the closed system is shown by the solid line for comparison. The horizontal line indicates the thermal expectation value at $T = 1$.}
    \label{fig:SM_weak_coupling}
\end{figure}

We now move on to numerical results for finite temperatures. Figure \ref{fig:SM_weak_coupling}(a) shows results for $\eta = 0.02$ as a weak coupling situation. When the temperature is sufficiently low, the residual energy decreases monotonically with ${t_{\rm a}}$. We expect that it should converge to the thermal expectation value, $\langle H_{\rm S}(1)\rangle^{\rm eq}_{{\rm I}, T_{\rm B}} + 1$, for $t_{\rm a}\to\infty$. For higher $T_{\rm B}$, the residual energy varies nonmonotonically with $t_{\rm a}$. This nonmonotonic curve can be divided in three regions: (i) the KZS for sufficiently small $t_{\rm a}$ where $E_{\rm res}$ decays following the power law as that of the closed system, (ii) the anti-KZS for the middle range of $t_{\rm a}$ where $E_{\rm res}$ deviates from the power law and increases with $t_{\rm a}$, and (iii) the quasistatic-freezing region for sufficiently large $t_{\rm a}$ where $E_{\rm res}$ decreases toward $\langle H_{\rm S}(1)\rangle^{\rm eq}_{{\rm I}, T_{\rm B}} + 1$. Note that the excess energy, defined by $\mathcal{E}_{\rm exc} \equiv \langle H_{\rm S}(1)\rangle_{t_{\rm a}} - \langle H_{\rm S}(1)\rangle^{\rm eq}_{I,T_{\rm B}}$ follows Eq.~(19) in (iii) as we mentioned in the main text. This nonmonotonic anti-KZS behavior of the residual energy has been discussed in literatures \cite{dutta2016, arceci2018}. It arises as a consequence of a large KZS region such that the residual energy at the end of the KZS region is below the thermal expectation value. 

\begin{figure}
    \centering
    \includegraphics[width=8cm]{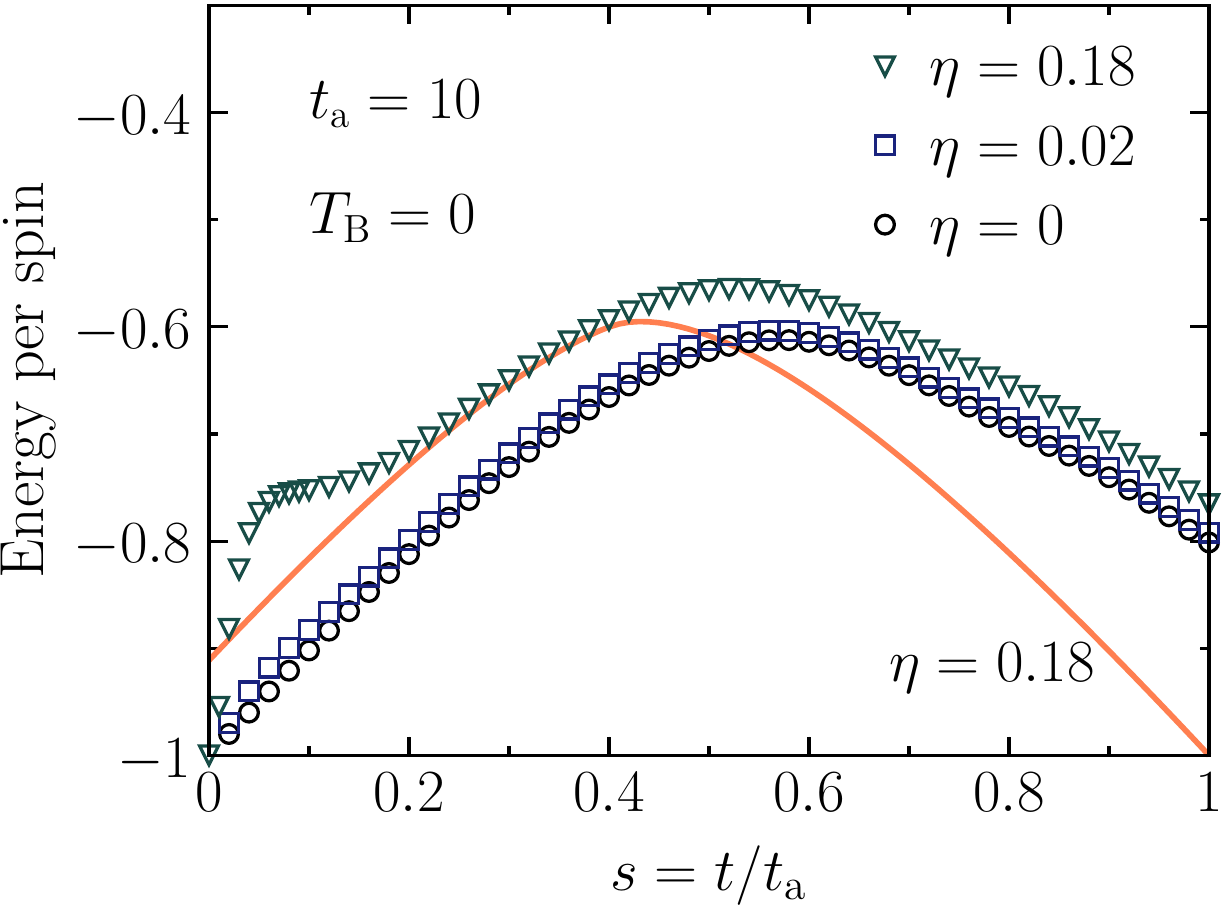}  \caption{Energies per spin as function of the rescaled time $s$ of time-dependent states with ${t_{\rm a}} = 10$  at $T_{\rm B} = 0$ and of equilibrium states at $T = 0$. Squares and triangles represent results of time-dependent states for ${\eta = 0.02}$ and $\eta = 0.18$, respectively, while circles are for the closed system. The solid line shows the instantaneous energies per spin of the dissipative system at $T = 0$ with $\eta = 0.18$.}
    \label{fig:SM_E_vs_s}
\end{figure}

\begin{figure}
    \centering
    \includegraphics[width=8cm]{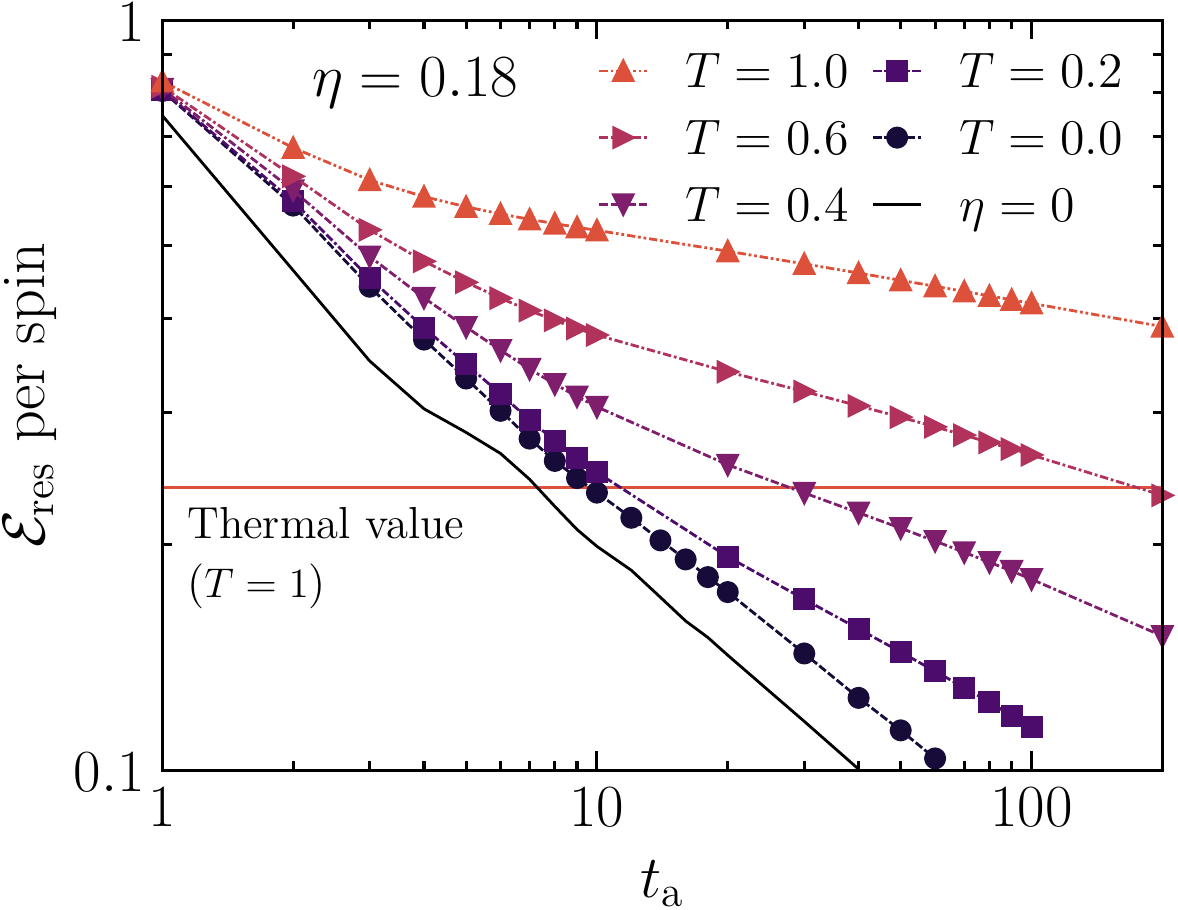}
    \caption{Residual energy after QA as a function of annealing time $t_{\rm a}$ for a medium coupling with $\eta = 0.18$ and various temperatures. The result of the closed system is shown by the solid line for comparison. The horizontal line indicates the thermal expectation value at $T = 1$.}
    \label{fig:SM_medium_coupling}
\end{figure}

We next focus on $\eta = 0.18$ as a medium coupling situation. Figure \ref{fig:SM_E_vs_s} compares the energy of a time-dependent state driven by QA with that of the instantaneous equilibrium state at $T = 0$. The spin system {equilibrates} with the bath in an early time of annealing, keeps equilibrium approximately for a while, and then drop out of equilibrium. The result for $\eta = 0.18$ is in contrast to the one for $\eta = 0.02$ where the energy of the time-dependent state is mostly indistinguishable from that in a closed system. 
The upward deviation of the curve of the time-dependent state at $T = 0$ from that of the equilibrium state comes from KZM, namely, the growth of the relaxation time near a quantum critical point and the failure of a quasistatic evolution. In Fig.~\ref{fig:SM_medium_coupling}, the residual energy at $T_{\rm B} = 0$ is observed to decay, following a power law with the exponent $b\approx 0.47$. As mentioned above, although this exponent is larger than $b_{\rm QMC} \approx 0.28$ predicted from the quantum Monte-Carlo simulation, a power law confirms KZS of the residual energy in a dissipative system. Raising temperature of the bath in Fig.~\ref{fig:SM_medium_coupling}, one finds that the range of ${t_{\rm a}}$ for KZS shrinks and eventually vanishes. The anti-KZS is not observed here. These results are in sharp contrast to the results for the weak coupling. At finite temperatures, QA does not cross a quantum critical point. The separation from a quantum critical point is larger and the relaxation time is shorter for higher temperature. When the relaxation time maximized near a quantum critical point at finite temperature is shorter than the time scale of QA, KZM does not work and a quasistatic evolution survives until freezing near the end of QA. This is why KZS disappears for higher temperatures. The absence of the anti-KZS is because the residual energy does not drop enough with KZS.

\bibliographystyle{apsrev4-1}
\bibliography{DTIM}
